\documentclass[a4paper,11pt]{article}
\pdfoutput=1
\usepackage{pos}
\usepackage[utf8]{inputenc}
\usepackage{textcomp}
\DeclareUnicodeCharacter{03BC}{\textmu}

\usepackage{amstext}

\usepackage[capitalise]{cleveref}

\usepackage[separate-uncertainty, exponent-product=\cdot]{siunitx}

\usepackage{nicefrac}
\usepackage{xspace}
\usepackage{subcaption}
\usepackage{commath}
\usepackage{placeins}
\usepackage{pgf}

\usepackage{wrapfig}

\newcommand{\cor}[1]{CORSIKA~{#1}\xspace}
\newcommand{\Xmax}{X_{\mathrm{max}}\xspace}
\newcommand{\sigmaXmax}{\sigma(X_{\mathrm{max}})\xspace}

\usepackage[backend=biber,
  articletitle=false,
  style=phys,
  giveninits=true,
  biblabel=brackets,
  chaptertitle=false,
  pageranges=false,
  doi=false,
  eprint=true,
  maxnames=2
]{biblatex}
\title{Avenues to new-physics searches in cosmic ray air showers}

\ShortTitle{Avenues to new-physics searches in cosmic ray air showers}

\author[a,b]{Oliver Fischer}
\author*[c,d]{Maximilian Reininghaus}
\author[c]{Ralf Ulrich}

\affiliation[a]{Max Planck Institute for Nuclear Physics, Heidelberg, Germany}
\affiliation[b]{Department of Mathematical Sciences, University of Liverpool, Liverpool, L69 7ZL, UK}
\affiliation[c]{Karlsruher Institut für Technologie (KIT), Institut für Astroteilchenphysik, Karlsruhe, Germany}
\affiliation[d]{Instituto de Tecnologías en Detección y Astropartículas (CNEA, CONICET, UNSAM), Buenos Aires, Argentina}

\emailAdd{oliver.fischer@liverpool.ac.uk}
\emailAdd{reininghaus@kit.edu}
\emailAdd{ralf.ulrich@kit.edu}

\abstract{
Cosmic Rays (CR) impinging on the terrestrial atmosphere provide a viable opportunity to study new physics in
hadron-nucleus collisions at energies covering many orders of magnitude, including a regime well beyond LHC energies.
The permanent flux of primary CR can be used to estimate event rates for a given type of new physics scenario.
As a step to estimate the potential for new-physics searches in CR-induced Extensive Air Showers (EAS), we here
determine the total luminosity, including the contribution stemming from the cascade of secondaries in hadron-air
interactions using Monte Carlo simulations of the hadronic shower component with \cor8. We show results obtained
for single showers and discuss the interplay with the CR spectrum.\\
Furthermore, we discuss the possibility to study BSM phenomenology in EAS, focusing on so-called large-multiplicity
Higgs production as an explicit example and its impact on EAS observables.
}

\FullConference{%
  40th International Conference on High Energy physics -- ICHEP2020\\
  July 28 -- August 6, 2020\\
  Prague, Czech Republic (virtual meeting)
}

\begin{document}
\maketitle

\section{Introduction}
Cosmic Rays enter the Earth's atmosphere with energies up to $\sim 10^{20}\,\si{\eV}$. They interact with air nuclei
at center-of-mass energies up to $\sqrt{s_{\mathrm{NN}}} \sim \SI{400}{\TeV}$ and therefore provide opportunities
to study elementary particle physics at energies far beyond LHC capabilities, however, with several obstacles: Due to the
steeply falling CR spectrum fluxes at the highest energies are tiny. For example, between $10^{17}\,\si{\eV}$ (corresp.\ to
$\sqrt{s_{\mathrm{NN}}} \simeq \SI{14}{\TeV}$) and $10^{19}\,\si{\eV}$ ($\sqrt{s_{\mathrm{NN}}} \simeq \SI{140}{\TeV}$)
the integral CR flux drops from \SI{5000}{km^{-2}.yr^{-1}.sr^{-1}} to \SI{0.3}{km^{-2}.yr^{-1}.sr^{-1}} in the H4a model~\cite{Gaisser:2011cc}.
Considering this, the common
way to study CR is indirect via the induced particle cascades consisting of billions of secondaries spreading over many
kilometers in the atmosphere, the so-called extensive air shower (EAS). Properties
of the primary CR are inferred from measured EAS observables.

Furthermore, the CR composition is a priori unknown and its indirect determination relies heavily on accurate models
of the hadronic interaction and extrapolations into phase-space hardly accessible in collider experiments. The
disagreement of composition measurements inferred from different EAS observables is commonly attributed to
a lack of understanding of the underlying hadronic physics, providing the opportunity to devise scenarios
of new physics.

In this contribution we aim to calculate the luminosity induced by the CR flux, including the contribution of secondary
interactions in the air showers. Additionally, we study the impact of \emph{large-multiplicity} events in the first interaction
on the EAS development, focusing on large-multiplicity Higgs production processes as particular example.

\section{Luminosity}
\begin{figure}[bt]
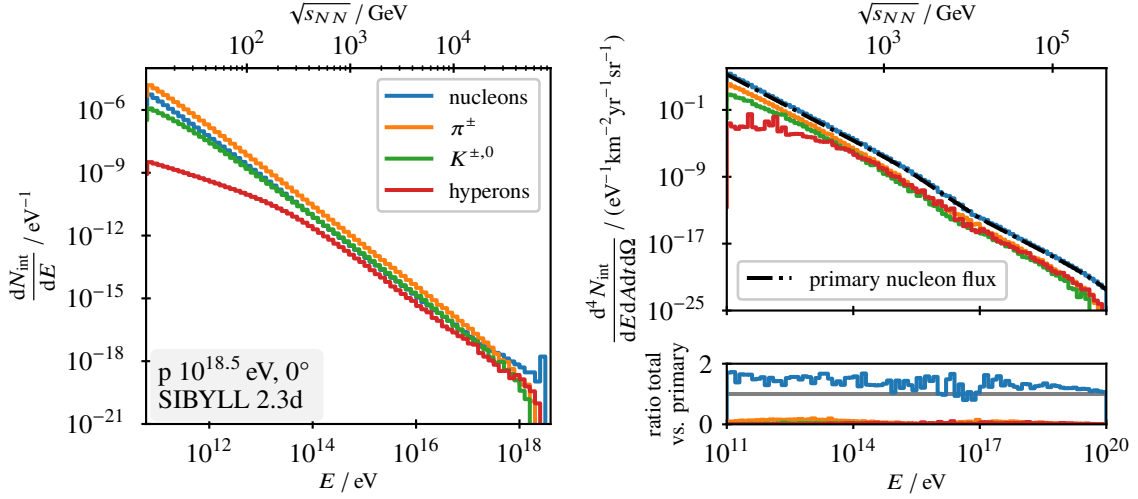

\begin{subfigure}[c]{0.5\textwidth}
%
%
\input{figures/dNintdE_18.5_0deg.pgf}\vspace{-1.5em}
\end{subfigure}
\begin{subfigure}[c]{0.5\textwidth}
%
%
\input{figures/inclusive_dNintdE_sib.pgf}\vspace{-1.5em}
\end{subfigure}
\caption{Number of hadronic interactions by species in a single EAS (left) as function of energy, and
folded with the CR spectrum (right)}
\label{fig:dNintdE}
\end{figure}

Using \cor8~\cite{Engel:2018akg} together with the hadronic interaction model SIBYLL~2.3d \cite{Engel:2019dsg,Ahn:2009wx}
we simulate the hadronic cascades of air showers in the energy range
$E_0 = \SI{100}{\GeV}$ to \SI{500}{\exa\eV}, with the energy spectrum and composition given by
the H4a model~\cite{Gaisser:2011cc} as implemented in the \texttt{crflux.models} package~\cite{crflux}.
In \cref{fig:dNintdE}~(left) we show the average number of hadronic interactions by species as a function of energy
for proton-induced EAS with energies of $E_0 = 10^{18.5}\,\si{eV}$. Due to the leading baryon
effect the "interaction spectrum" close to the primary energy is dominated by nucleon-air (anti-nucleons included) interactions,
with the peak at the upper end given by the primary interaction. Only at energies about an
order of magnitude lower, pion-air interactions start to overtake. Together with the other species they
display a power-law behaviour (roughly $~E^{-2}$) for a large energy range until decays start to become relevant.

In contrast, inclusive fluxes are dominated by nucleon interactions over the whole energy range, cf.~\cref{fig:dNintdE}~(right).
This results from the fact that the primary CR spectrum is steeper than the interaction spectrum of single EAS and only showers with energies $E_0$ slightly greater than $E$ contribute significantly to $\mathrm{d}N_{\mathrm{int}} / \mathrm{d}E$.
All in all, the total flux including the secondary flux is greater than the primary nucleon flux by less than a factor of two. 
Our results are comparable to those obtained in ref.~\cite{Illana:2006xg}, where a simple power-law primary nucleon flux was assumed.

\begin{figure}
\centering
\input{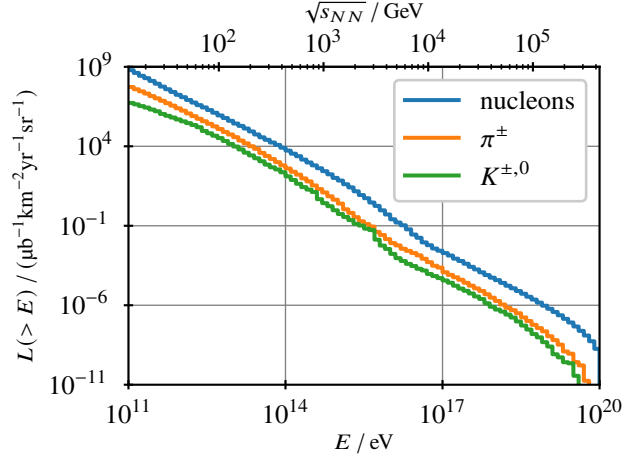}\vspace{-1.5em}
\caption{Integral total (primary plus secondary) luminosity of nucleons, pions, and kaons from CR.}
\label{fig:inclusiveLumi}
\end{figure}

In \cref{fig:inclusiveLumi} we convert these numbers to luminosity to be more useful in comparisons with
accelerator experiments. We define
\begin{equation}
	L(>E) = \int_E^{\infty} \frac{\mathrm{d}N_{\mathrm{int}}}{\mathrm{d}E'} \frac{1}{\sigma_{\mathrm{prod}}(E')} \, \mathrm{d}E',
\end{equation}
where the production cross-sections $\sigma_{\mathrm{prod}}$ for the individual species are those provided by
the hadronic interaction model. For example, integrated over the whole surface of the Earth, the luminosity
above $\sqrt{s_{\mathrm{NN}}} = \SI{10}{\TeV}$ is about \SI{22}{pb^{-1}.yr^{-1}}. Above \SI{100}{\TeV} 
it is still about \SI{8}{\micro b.yr^{-1}}.

\section{Large-multiplicity scattering}
Scattering processes with large final state multiplicity at very high energies have been studied for instance for the SM \cite{Goldberg:1990qk}.
The amplitude of such processes, which have final states with $n\gg 1$ Higgs bosons, grows factorially with $n$ due to the large number of contributing Feynman diagrams \cite{Cornwall:1990hh,Goldberg:1990qk}.
Provided that this growth of the amplitude reflects reality (rather than an incomplete calculation) and under the assumption that perturbative unitarity is ensured \cite{Khoze:2017tjt}, the cross section must be bounded from above, favouring a specific large process energy and thus number of Higgs bosons produced. Taking Fig.~6 of ref.~\cite{Degrande:2016oan}
at face value, the energy fraction of proton-proton collisions at $\sqrt s \geq \SI{50}{\TeV}$ converted into Higgs bosons can reach $\mathcal O(\SI{30}{\percent})$.

\subsection{Modelling $n$-Higgs production in air showers}
We include the large-multiplicity process phenomenologically into our CR simulations to quantify observable features in the EAS spectra.
We simulate proton-induced, vertical showers of the primary energy $E_0 = 10^{19}\,\si{\eV}$. Since the proton-nucleus cross-section, which determines the distribution of the height of the primary interaction $h_0$, of a large-multiplicity process is unknown, we set a fixed $h_0 = \SI{20}{km}$ above sea level.

We model the primary event in the following way: A fixed
fraction $f$ of the center-of-mass-energy per nucleon $\sqrt{s_{\mathrm{NN}}} = \sqrt{2 m_{\mathrm{N}} E_0}$ is used to create Higgs particles,
while the underlying event is generated with QGSJetII-04~\cite{Ostapchenko:2010vb} with the remaining energy $(1-f)\times E_0$.\footnote{
Strictly speaking, this procedure is only an approximation since a fraction of the actual phase space
stays unavailable for hadron production in QGSJetII in our implementation. This could be alleviated by generating the event
with the whole energy and replacing a fraction of the secondaries by Higgs particles afterwards.}
A second parameter $\varepsilon$ is introduced to model the kinetic energy of the Higgs particles relative to their mass.
To ensure momentum conservation we always generate an even number $n_{\mathrm{h}}$ of Higgs particles in $n_{\mathrm{pairs}} = n_{\mathrm{h}} / 2$
pairs with their momentum vectors back-to-back. The number of pairs is given by
\begin{equation}
n_{\mathrm{pairs}} = \left\lfloor \frac{f \sqrt{s_{\mathrm{NN}}}}{2 m_{\mathrm{h}} (1 + \varepsilon)} \right\rfloor,
\end{equation}
where we consider $\varepsilon =0.1,\,1,\,2$ for concreteness.
The available total kinetic energy $T = f \sqrt{s_{\mathrm{NN}}} - n_{\mathrm{h}} m_{\mathrm{h}}$ is split among the pairs by distributing the individual energy portions per pair $T_{\mathrm{pair}}^{(i)}$
uniformly on the $(n_{\mathrm{pairs}} - 1)$-simplex defined by the conditions
\begin{equation}
\sum_{i=1}^{n_{\mathrm{pairs}}} T_{\mathrm{pair}}^{(i)} = T \quad \text{and} \quad 0 \leq T_{\mathrm{pair}}^{(i)} \leq T.
\end{equation}
We distribute the momentum vectors isotropically in the proton-nucleon center-of-mass system.
In the subsequent course of the simulation we let the Higgs particles decay (using Pythia~v8.235~\cite{Sjostrand:2014zea})
into long-lived particles. Electromagnetic (EM) particles occuring during the simulation are fed into CONEX~\cite{Bergmann:2006yz},
which will generate EM longitudinal profiles by solving the cascade equations numerically. Secondary hadronic interactions are treated
with QGSJetII-04 above and with UrQMD~1.3~\cite{Bleicher:1999xi} below \SI{60}{\GeV}.

\subsection{Results}
In this section we study the impact of large-multiplicity events on the shower maximum as well as the total number and the energy spectrum of muons at ground.

We define the shower maximum $\Xmax$ as the position along the shower axis (measured in slant
depth $X$) at which the electromagnetic energy deposit by ionization and absorption of
low-energetic electrons is maximal. It is a quantity that is routinely measured with fluorescence
detectors on an event-by-event basis. The first two moments, mean $\langle\Xmax\rangle$ and
standard deviation $\sigmaXmax$, of its distribution are commonly employed to determine the
mass composition of UHECR~\cite{Kampert:2012mx}.

\begin{figure}[bt]
\centering
\input{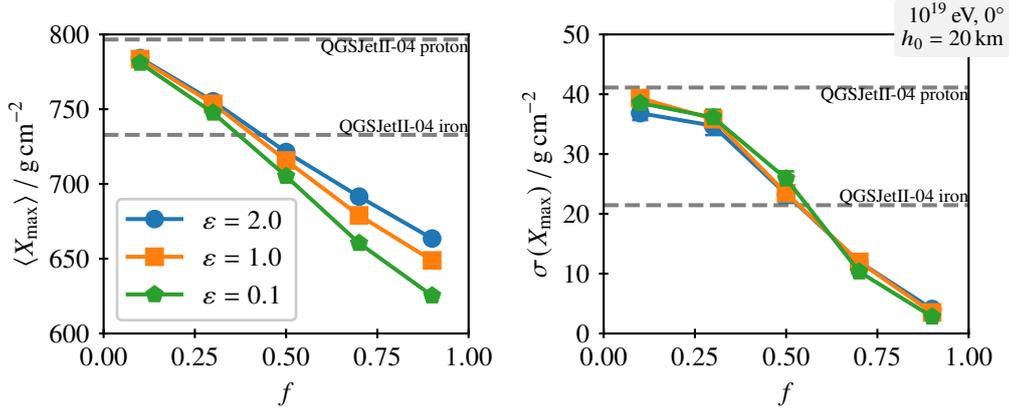}\vspace{-1.5em}
\caption{Dependence of the two moments of the $\Xmax$ distribution on $f$ and $\varepsilon$ with fixed $h_0$}
\label{fig:xmax}
\end{figure}

In \cref{fig:xmax} $\langle\Xmax\rangle$ and $\sigmaXmax$ are shown as a function of $f$ and $\varepsilon$. $\langle\Xmax\rangle$ decreases almost linearly with $f$ with a slope only mildly depending on $\varepsilon$. Also the fluctuations $\sigmaXmax$ are reduced with increasing $f$, almost independent of $\varepsilon$. Furthermore, the values obtained for proton and iron showers without new physics are indicated in \cref{fig:xmax} as dashed lines. 
Consequently, disregarding the artificially fixed $h_0$, for some values of $f$ and $\epsilon$ the large-multiplicity showers could be incorrectly classified as due to heavier primaries if interpreted using only standard hadronic interactions. It appears that values of $f \gtrsim \num{.4}$ are disfavoured by the fact that such showers
seem heavier than iron.

%
%

\begin{figure}[bt]
\centering
	\input{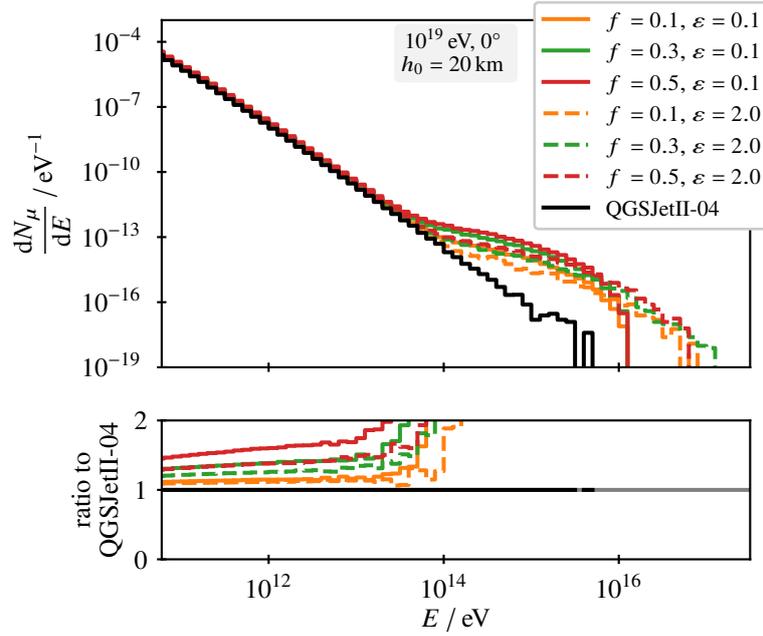}\vspace{-1.5em}
\caption{Secondary muon energy spectrum at sea level (\SI{1036}{g.cm^{-2}})}
\label{fig:muspect}
\end{figure}

The energy spectrum of muons at ground is depicted in \cref{fig:muspect}. The most striking
feature is the increase of muons with energies above $\gtrsim \SI{10}{TeV}$. These are prompt
muons stemming from the decay of the Higgs bosons. Their maximum energy is shown to be dependent on $\varepsilon$.
Moreover, we observe an increase up to \SI{50}{\percent} depending on the model parameters throughout the whole energy spectrum with only a slight change in the overall slope.

We notice that the here presented muon spectra are qualitatively comparable to those obtained by~\textcite{Brooijmans:2016lfv} with sphalerons as first interaction, suggesting a somewhat general signature of electroweak physics within the first interaction in EAS.

%
%

\section{Conclusions}
We have shown that the luminosity induced by primary and secondary CR interactions is mainly determined by
primary nucleons, with secondary nucleons increasing this number by up to $\sim\SI{80}{\percent}$. Though
not competitive with LHC luminosity they are a unique laboratory to study particle physics at
energies much greater than the LHC. As a particular example of new physics we have demonstrated that
large-multiplicity Higgs production events have a significant impact on the distribution of the shower
maximum and the muon energy spectrum.

\section*{Acknowledgements}
M.R. acknowledges support by the Doctoral School KSETA.
The simulations were performed on the bwForCluster BinAC of the University of Tübingen, supported by the state of Baden-Württemberg through bwHPC and the DFG through grant no. INST~37/935-1~FUGG.



\printbibliography

\end{document}